\documentclass[manuscript,nonacm]{acmart}
\renewcommand{\authorsaddresses}{}

\makeatletter
\renewcommand{\@authorsaddresses}{}
\makeatother

\setcopyright{none}
\renewcommand\footnotetextcopyrightpermission[1]{}

\AtBeginDocument{%
  }

\acmConference{}{}{}
\acmBooktitle{}
\acmYear{2026}
\copyrightyear{2026}
\acmDOI{}
\acmISBN{}

\usepackage{xcolor}
\usepackage{multirow}
\usepackage[normalem]{ulem}

\setcopyright{acmlicensed}

\author{Victoria Chui\textsuperscript{*}}
\author{Kelly McConvey}
\author{Daniel Chui}
\author{Malayna Bernstein}
\author{Shion Guha}

\affiliation{
  \institution{University of Toronto}
  \country{Canada}
}

\begin{document}

\title{Embedded Human-Centered Data Science in a Graduate Programming Course: A Framework and Case Study}

\renewcommand{\shortauthors}{Chui et al.}

\begin{abstract}
As AI and data-driven systems pervade practice, there is an imperative for instructors to embed societal impact and ethics content into computing courses. In response, we present the Human-Centered Education for Learning in Information and eXplainable Computing (HELIX) framework for information science programs, organized around three iterative pillars —knowledge building, decision-making, and empowerment —with concrete actions for instructors and students. We applied the framework in a graduate, introductory programming course using readings, algorithmic design activities, and scenario-based reflections. 
We present a pilot implementation of this framework to examine changes in students' (n=22) knowledge acquisition, decision-making processes, and self-reflection regarding human-centered perspectives in data science.
We release an anonymized materials kit (survey, assignments, analysis code) to support adoption. We discuss design tensions (workload, assessment, relevance to diverse information science learners) and provide guidelines for integrating human-centered content without overwhelming technical outcomes. Findings suggest that the HELIX Framework is feasible in information science contexts and future work should use comparative survey assessment to strengthen causal inferences.

\end{abstract}





\maketitle

\section{Introduction}

With the increase in computational power and potential for technological tools in workplace decision-making, responsible and ethical algorithmic design choices must be progressively prioritized. Human-centered methods such as responsible and justice-centered computing enable fairer outcomes from data-driven solutions \cite{baumer_HCAD}. This imperative has yielded a growing interest in introducing algorithmic bias, justice, and ethics content in university-level computing courses. To ensure that future generations of computational experts are aware of these needs, researcher-educators have begun embedding human-centered content and real-world examples in computer science and related courses \cite{horton_embedded_2024,horton_embedding_2022,horton_is_2023,fiesler_integrating_2021,cs4sg}. Researchers have synthesized the main takeaways from this content and raised awareness of the major barriers to widespread adoption of integrating ethics in the classroom \cite{brown_teaching_2024}. These challenges include a lack of cohesion with existing technical course learning outcomes, the addition of content leading to overwhelming course loads, and the necessary increase of preparation for instructors \cite{brown_teaching_2024}. 

In response to these imperatives and associated instructional challenges, we propose and preliminarily evaluate a three-pillared educational framework for embedding human-centered topics into technical, data science courses in information science (IS) departments. This Human-Centered Education for Learning in Information and eXplainable Computing (HELIX) Framework is designed for IS students throughout their data science degrees, inspired by - but distinct from - computer science ethics initiatives due to IS students' wide ranging educational backgrounds (often less technically-focused than computer science) and broad degree-level learning outcomes. These learning outcomes centre around topics such as societal impact, social responsibility, intellectual growth, and critical assessment. The pillars of the framework mirror students' learning stages within a single course and across their degree, from knowledge building (understanding definitions and concepts) to decision making (applying their human-centered knowledge to technical solutions) to empowerment (taking these learning outcomes forward). This work situates ethics within a broader human-centered computing perspective that includes social justice concerns such as equity, fairness, and accountability. We developed this framework by adapting existing computing ethics pedagogy for data science contexts and embedding relevant human-centered content into IS computing courses, and include a case study to illustrate and preliminarily evaluate the efficacy of the framework.

While computing education has predominantly been taught in computer science departments, the ubiquitous relevance of computing across many other fields had led to other departments developing and instructing their own computing courses \cite{cooper}. This is prevalent in `professional programs' which focus heavily on preparing (often graduate) students for their future careers. Computing courses in these programs are often taken by students from a range of backgrounds, which may not include computer science or programming. The field of data science has grown out of computer science and statistics, and is a prevalent stream within professional programs. Data science and human-centered courses have followed the natural evolution of increased computational and Big Data potential and the need to prepare a range of students for data-driven tasks in the workplace \cite{saltz_integrating_2019,Donoho}. Computing education scholars and instructors now need to provide students with a broader range of technical competencies \cite{Hedayati}. Students from information and data science programs apply their technical skills to practical applications, often learning through hands-on case studies, paralleling the professional nature of their degrees \cite{cao}.

Research on embedding human-centered and ethical content requires further validation in computing courses taught at IS departments - distinct from computer science courses due to the interdisciplinary, professional nature of IS degrees and their focus on human interactions with algorithms and data. Programming courses in the IS field teach students technical skills in languages such as Python and R, alongside the tools for analyzing the impacts of algorithmic systems on society. Due to the multidisciplinary nature of IS degrees, the program learning outcomes (PLOs) can be broad and incorporate multiple theoretical frameworks. In the \textcolor{red}{redacted department at the institution}, the PLOs include a dedication to information theories and practices, cultural and social leadership, critical assessment of information, and societal implications of technical developments \textcolor{red}{[ANON]}. 

To encourage further integration of the multidisciplinary PLOs into a data science stream within an IS department, we developed the HELIX Framework. Rather than replacing existing computing ethics pedagogy, the HELIX Framework adapts these approaches to the interdisciplinary context of IS and data science education. We define human-centered data science (HCDS) as the coalescence of human, user, and societal considerations with the development of technical tools through computing. This definition follows from previous studies exploring data-driven, HCDS techniques in fields such as health, education, housing, social welfare, and finance \cite{McConvey_Guha_Kuzminykh_2023,Moon_Guha_2024,saxena_lit,pam_lit,chui_compass,Chui_Pater_Toscos_Guha_2023}. Our framework (Figure \ref{fig:framework}) reflects the stages of student learning throughout a course, as well as actions for the student and instructor, and common challenges. It outlines strategies for embedding content on human-centered algorithm design, human values, and critical theory. We then preliminarily tested this framework by integrating such content into the instruction and assessment of an introductory programming course during the \textcolor{red}{redacted semester} to further understand the development of students' leadership skills, knowledge of information theory and practices, ability to critique algorithmic design, and motivation to think critically in future workplace settings. 
Students' understanding and knowledge of this content was assessed with pre- and post-course surveys. The full scope of the research--the framework development, its integration into the course curriculum, and its adoption by students--was designed to address the following research questions:

\begin{itemize}
    \item \textbf{RQ1}: What components (pillars, actions, learning outcomes, challenges) emerge for a Human-Centered Education for Learning in Information and eXplainable Computing (HELIX) framework from a structured synthesis of prior literature and IS program learning outcomes?
    \item \textbf{RQ2}: Among master’s students in an introductory programming course where the HELIX Framework is embedded, how do the following change from pre to post?
    \begin{enumerate}
        \item \textbf{(Knowledge):} knowledge of human-centered data science concepts. 
        \item \textbf{(Attitudes/Perceptions):} attitudes toward ethics, privacy, fairness, and societal impact. 
        \item \textbf{(Decision Making):} selections in ethically-oriented applied scenarios. 
        \item (\textbf{Program Learning Outcomes}): students’ perceived advancement of IS PLOs.
        
    \end{enumerate}
\end{itemize}

Building on previous researchers' experiences embedding ethical content in undergraduate computer science courses \cite{horton_embedded_2024,horton_embedding_2022,horton_is_2023,fiesler_integrating_2021}, we introduced a framework to accommodate information and data theories, practices, and human-centered content in graduate IS programming courses. We applied our framework to a single course case study, embedding content in a semester-long, introductory programming course in an IS department. Quantitative analysis of our pre- and post-course surveys showed an overall significant difference in students' performance on the survey (further described in \ref{sec:method_quant}) with p = 0.001. The attitudes and perceptions portion of the survey, composed of five Likert questions, also had a significant difference (pre-course mean of 20.8/25, post-course mean of 22.7/25, p = 0.002). 
Lastly, students' self-reporting of improved understanding regarding HCDS indicated that 21 of our 22 respondents saw `somewhat' or `significant' improvement.

Our work aims to further analyze the potential for human-centered activities in IS programming, covering a wider range of topics and student backgrounds than a typical introductory computer science (CS1) course. We contribute the following to future computing education and research:

\begin{itemize}
    \item We performed a literature scan of existing PLOs for data science students in IS programs and education frameworks on student learning. We synthesized these insights into a three-pillared framework for human-centered data science computing education (HELIX Framework), populating this framework with actions for both students and teachers, alongside collaborative goals at each pillar, and potential challenges.
    \item We conducted a single, preliminary case study, applying our framework to graduate-level introductory programming course in an IS department. We performed a quantitative analysis of students' pre- and post-course survey scores, examining performance trends across subsets of the framework ({\ref{sec:quant}}), and paid specific attention to the PLOs that students anticipated learning (pre-course) and those that students felt they did advance (post-course). 
    \item We present the components of our HELIX Framework for an IS department. We outline future steps and guidelines for introducing this content into IS programming courses in a diverse, multidisciplinary faculty producing graduates who will work in a broad range of jobs (\ref{sec:guidelines}). This content is provided as an open \textcolor{red}{(currently private for submission)} resource for researchers to examine, test, adapt, and refine the HELIX Materials Kit v1.0, \ref{sec:kit}), including our survey, a synthetic dataset, replicable quantitative code, and assignment adaptations \cite{zenodo_id}.
\end{itemize}

\section{Background}



With the rise of artificial intelligence (AI) and algorithmic, data-driven solutions in society and workplaces, researchers are emphasizing the need for human-centered instruction for future programmers, engineers, and computer science experts \cite{goetze_integrating_2023}. This is paralleled in the updated ACM Code of Ethics and Professional Conduct, with specific reference to the education of future programmers \cite{ACM_ETHICS}. Researchers have also emphasized the need for a wider range of ethical considerations in relevant frameworks, towards an increasing number of ethics lessons in computing education \cite{saltz_integrating_2019}. Horton et al. prompt educators to empower their computing students to conduct ethical decision-making in the workplace \cite{horton_embedded_2024}. They additionally emphasize the need to prepare students to anticipate, mitigate, and take responsibility for the societal impacts of their technologies \cite{horton_is_2023}. Research efforts have also focused on justice-centered programming instruction, with Kivuva et al. integrating content on AI applications, bias and justice, and societal impact, with the goal of empowering students to think critically about the impacts of AI, towards social justice \cite{Kivuva}. A push towards interdisciplinary education in computing highlights the need to complement technical instruction with philosophical and theoretical concepts \cite{goetze_integrating_2023}. In addition, previous programming education has not always included concerns around diversity and inclusion, for either the students receiving this education, or those impacted by the solutions being developed in class. The desirability of ethical content embedded in programming courses is driven by the ``boom" of large language models, AI, and algorithmic implementations in work and social lives. By prompting students to engage with these concepts and subsequent societal concerns raised by these technologies during their early programming courses, a strengthened interdisciplinary education can be derived \cite{horton_embedded_2024,horton_is_2023,goetze_integrating_2023}.





Multiple challenges have been previously raised when embedding ethical topics into computing courses. These challenges include an increased workload due to additional introduced content, time and labour strains for teaching assistants (TAs) and instructors, and a lack of understanding regarding how applicable ethics is in the workplace. Fiesler et al. observe that when integrating ethics into a collection of introductory programming courses, students (especially those from non-technical academic backgrounds) would often prioritize the technical concepts over ethical content (the latter of which was not graded) \cite{fiesler_integrating_2021}. Researchers recommend that additional ethics content must be integrated in such a way that it does not strain or overwhelm pre-existing course learning goals in order to remain engaging for students \cite{petelka_principles_2022,smith_incorporating_2023}. Additionally, teaching staff take on additional labour due to increased content being integrated on a weekly basis, particularly in interdisciplinary courses. Students and instructors have both shown a desire to include tech experts and ethicists in course instruction and design \cite{skirpan_ethics_2018,smith_incorporating_2023,ahuja_understanding_2024,tran_its_2024,saltz_integrating_2019}. Surveys and interviews from such studies suggest that student engagement is furthered with guest lecturers that can speak in detail to relevant ethical considerations and workplace stakeholders. Lastly, major concerns around the sustainability of the ethical learning outcomes in future study and workplace settings have been raised. Researchers emphasize the need to include real world impacts in embedded content and prompt students to discover and mitigate their own societal impacts. Encouraging students to ask ``how to think, not what to think" \cite{horton_embedding_2022} can further integrate their ethical training in practice. Maintaining this embedded content in subsequent computing courses is an actionable way to continue students' learning of both technical and theoretical programming concepts, enriching their comprehension of the societal impacts of their work across their degree \cite{horton_is_2023}. Horton et al.'s [2024] survey concluded that those students that could recognize ethical concerns in the workplace were more likely to have taken a previous course with embedded ethics; however, they were no more likely to feel empowered to resolve the societal impact conflict \cite{horton_embedded_2024}.

HCI has a long history of adaptability to the programming applications prevalent in society \cite{history_HCI}. The field investigates the interactions between technologies, programs, and applications with the relevant users \cite{history_HCI}. It has consistently been promoted as an interdisciplinary subject, combining education from ``psychology, computer science, design, anthropology, information science, and others" \cite{history_HCI}, emphasizing cross-departmental, cross-field instruction. In some educational settings, HCI content is integrated into programs centered on user experience design (UXD). There also exists an intersection of critical theory with HCI, with researchers encourage the integration of critical theory concepts with the development of technical tools \cite{critical_theory_CS,obrist}. This intersection prompts human-centered considerations, yielding Baumer's Human-Centered Algorithm Design (HCAD) framework \cite{baumer_HCAD}. The HCAD framework has been extensively studied across multiple fields, prompting researchers and programmers to anticipate and mitigate the societal impacts of their work - including AI models and tools \cite{chui_compass,Chui_Pater_Toscos_Guha_2023,Moon_Guha_2024,McConvey_Guha_Kuzminykh_2023}. Applying human-centered considerations to data science work and education has been an increasing call in the ACM SIGCHI and wider computing communities, particularly with the rise of generative AI \cite{hcds_book,critical_ds_workshop,gen_AI}. 

Literature on embedded ethical content most frequently considers computer science, undergraduate courses. While these challenges and takeaways are impactful for instructors across multiple fields, there is a need to understand the unique challenges and possibilities of embedding human-centered content in graduate IS courses. These students often come from a wide range of (non-technical) backgrounds, and have broad interests - from library science and knowledge management, to user experience design and data science. Understanding how computing is taught in this field, and the potential for human-centered content, will yield a stronger generation of IS scholars in the workforce.

\section{Methods}

\subsection{Framework Development} \label{sec:framework}

\begin{figure}[t]
    \centering
    \includegraphics[width=1\linewidth]{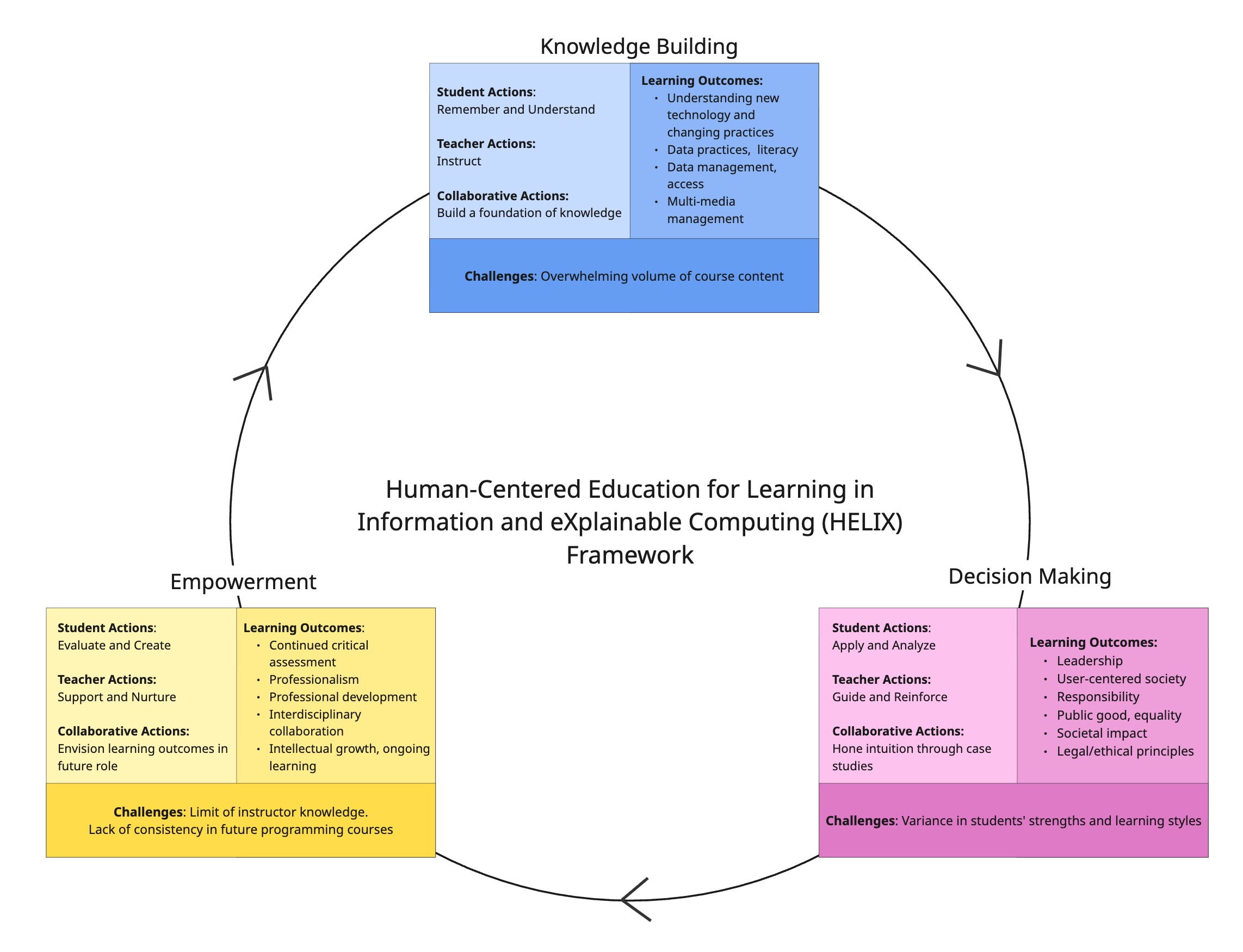}
    \caption{HELIX Framework}
    \label{fig:framework}
\end{figure}

Through a literature search, the HELIX Framework was developed around three foundational pillars: 1) knowledge building, 2) decision making, and 3) empowerment (Figure \ref{fig:framework}). The HELIX Framework was designed to complement existing computing ethics curricula, by adapting concepts from existing  curricular frameworks and learning outcomes from multiple IS departments, in order to be effective in interdisciplinary data science contexts. The literature search had multiple components, which we explain in detail before discussing the structure of the final framework:
\begin{enumerate}
    \item Frameworks in educational settings
    \item PLOs across IS programs with data science instruction
    \item Challenges in embedded ethics in computing courses
\end{enumerate}

\subsubsection{Frameworks in Education} \label{sec:edu_frame}

Previous frameworks in education were explored to inform the iterative, three-pronged structure of the HELIX Framework. To find academic frameworks on computing and education, the following databases were systematically search: the ACM Digital Library, IEEE Xplore, Google Scholar, and Elsevier with the following keywords: ``educational framework,'' ``curricular framework,'' ``student learning journey,'' ``ethics education framework'', ``computing ethics pedagogy,'' and ``computing education.'' We began by examining the stages presented in the broadly-applied Bloom's Revised Taxonomy \cite{bloom_revised} before realizing we needed to more intricately engage with prior literature on critical pedagogies \cite{mayhew_patitsas}, the interaction between social, ethical, and technical knowledge units \cite{martin_huff}, and AI literacy and computational competencies \cite{long,unesco}. We found additional frameworks through works cited within these preliminary papers. To narrow down the most relevant, we removed any papers that specialized solely in primary or secondary education (those that didn't mention post-secondary). We also removed papers that didn't outline a structured pedagogical framework or those that were not peer-reviewed. We did not restrict our search to only those frameworks in computing education, but included those more broadly applicable to post-secondary education. Each framework was analyzed to identify recurring pedagogical components such as knowledge acquisition, applied reasoning, and student agency. Terms from these frameworks informed the naming of the three pillars of our HELIX Framework. Components from these frameworks informed the iterative, pillar structure of the HELIX Framework,  which emphasizes foundational knowledge, applied decision-making, and learner empowerment (Figure \ref{fig:framework}). 
The HELIX Framework reflects the iterative nature of learning with a developmental focus that includes knowledge development of broad concepts, the transfer of those concepts to the applied contexts of code development and design, and the promotion of critical evaluation and empowerment.
We synthesized three `pillars,' using a cyclical structure to encourage iterative and flexible movement during learning. As students' learning is often not linear, and requires revisiting concepts, the cyclical structure of the framework encourages both students and teachers to transition between the pillars fluidly and repeatedly. The three-pillar structure parallels progression models such as Bloom's taxonomy; however, the framework adapts this progression for HCDS by emphasizing responsible decision-making and student agency within sociotechincal systems.

The frameworks that were most useful in informing the structure of the HELIX Framework were those focused on HCI/computing education, post-secondary education, and computational competencies.
We synthesize tenets from multiple of these curricular frameworks to derive the `Actions' portion of each pillar of our framework. Bloom's Revised Taxonomy preliminarily inspired the inclusion of `student actions' at each pillar, with the precise actions for students and teachers being informed by computing education frameworks such as \cite{soh_integrated_2007}. We also encourage an active learning approach, which researchers have previously recommended in data science education, as it can increase students' autonomy over their learning and confidence in their technical abilities \cite{active_learning_1,sanusi}. The role of the instructor at each pillar was informed by \cite{martin_huff}, from a teacher, to a guider, to a supporter, assisting active student learning throughout. We also highlight collaborative actions for students and teachers together, providing a common goal at each pillar (Figure \ref{fig:framework}). These actions are driven by takeaways from prior precise curricular frameworks, such as critical pedagogies, encouraging students and instructors to work together to understand power structures in computing classrooms \cite{mayhew_patitsas}. Martin and Huff's framework on the interaction of social, ethical, and technical knowledge informs these collaborative goals, encouraging students and teachers to consider social, ethical, and technical concepts. Two frameworks on AI literacy and competencies encourage reflection of critical thinking, transparency, the human role, critical interpretations of data, and interdisciplinary collaboration \cite{unesco,long}. While these frameworks are intended for education in AI, we can apply several concepts to more general computing courses.



\subsubsection{Program Learning Outcomes}

The iSchools is an organization of 134 IS departments, providing member institutions with access to grants, the iConference, communities, and regional opportunities to connect. The members of The iSchools are united in their commitment to education and research in the broad field of information. Program Learning Outcomes (PLOs) were used as a proxy for institutional priorities in the absence of consistent access to syllabi across institutions. As PLOs provide a non-exhaustive listing of course or program content, future work should examine course syllabi, accreditation criteria, and student/instructor perspectives to better characterize how human-centered concepts are implemented. We used the iSchools Members Database \cite{ischools} to explore PLOs from participating institutions around the world. Each institution that has an active website, and provides instruction in data science topics (as a stream, degree, or course topic) was explored for PLOs. To identify such institutions, we examined each of the 134 entries in the iSchools Members Database that had an active website on their profile. From the website, we identified if data science, AI, and/or machine learning is offered as a topic, course, stream and/or program. If data science instruction is present, we then examined the website for PLOs. Some institutions had an identifiable and labeled set of learning outcomes; however, many only had such outcomes embedded in ``What Our Students Learn'' or similar subsections. These latter outcomes were not included in our analysis since they were often promotionally phrased and not labeled intentionally as `Learning Outcomes.' 

Analyzing PLOs provided insights into how institutions conceptualize the role of ethics, social justice, and human-centered instruction within IS curricula. We collected PLOs from over 50 institutions and conducted iterative, inductive thematic analysis \cite{braun} to identify higher-level themes. Preliminarily, 1-4 codes were attributed to each learning outcome. For example, for the following outcome, `\textit{Integrate concepts from information/data management, digital technologies and
human behavioral and cultural practices to help solve organizational, community
or social challenges.'} \cite{UWISC} we assigned preliminary codes including \textit{data management, technologies, cultural behaviour,} and \textit{social impact}. Iterating through each of the identified learning outcomes, we assigned a total of 27 preliminary codes. These codes were then grouped into themes. For example, information practices, changing practices, information management, and information literacy codes were grouped into `Understanding Information.' Similarly, public good, social impact, responsibility, leadership, and user-centered society were grouped into `Human-Centered Decisions.' Our last major theme of `Future Empowerment' including codes such as professional conduct, intellectual growth, life-long learning, and collaboration. Two of the co-authors iteratively discussed this `theme-building' process through a consensus building approach. After each stage of this PLO extraction process (collecting PLOs, assigning codes, grouping into themes), joint discussion resolved any uncertain nuances. Further codes for each of these themes are outlined below:
\begin{enumerate}
    \item Understanding information practices, new technology, information literacy, information access, changing practices, multi-media management, data management
    \item Leadership, user-centered society, responsibility, public good, equality, societal impact, legal/ethical principles
    \item Continued critical assessment, professionalism, professional development, interdisciplinary work and collaboration, intellectual growth, ongoing learning, advocating for responsible design
\end{enumerate}

We then connected these three themes to the actions outlined in \ref{sec:edu_frame}. Since the first theme was related to understanding information practices and management, we connected it to the Remember and Understand portion of the framework, where students and teachers strive to collaboratively build a knowledge foundation in these concepts. This set of PLOs relate to understanding and remembering data practices and technological developments, alongside data management. 

The second set of PLOs focus on the role of the student when applying technology to society, with concepts that guide students' decision-making: societal impact, ethical principles, and social leadership. We connected this PLO theme to the Apply and Analyze actions, where the collaborative goal is to simultaneously advance students' computing intuition while integrating the impacts of their work on society. 

Finally, many iSchools had at least one PLO related to continued intellectual growth and/or critical assessment in the workplace. This set of PLOs reflects the ``professional'' nature of information science degrees, particularly at the graduate level, aiming to prepare students for their future careers and encourage ongoing learning. With respect to our framework, Empowerment additionally refers to students' perceived capacity to question, critique, and intervene in the development of data-driven systems. We attributed this final PLO theme to the Evaluate and Create set of student actions, with the collaborative goal of envision these learning outcomes in subsequent education or workplace settings. This component of the framework looks forwards to students' future courses or professional roles, empowering them to continue making human-centered, reflexive decisions.

\subsubsection{Challenges in Embedded Ethics} \label{sec:challenges}

Previous research on embedding ethics in computing courses (in computer science, data science, and/or information science) was examined to synthesize the greatest challenges for instructors and students. To search for relevant literature we used keywords such as ``embedded ethics,'' ``integrated ethics,'' ``human-centered computing,'' ``ethics in HCI,'' and ``ethics in CS1,'' in the ACM Digital Library and IEEE Xplore, with particular attention to ACM Transactions on Computing Education (TOCE), CHI, ICER, and SIGCSE venues. For a paper to be included in our search it had to involve at least one application or study of embedding ethics/human-centered content in a computing or HCI course, with noted challenges for instructors and/or students. We also examined the works cited by each identified paper to further discover relevant studies. Brown et al.'s literature review of ethics in computing courses highlights five major challenges, supported by other SIGCHI literature on embedding ethics and human values in computing/HCI courses \cite{brown_teaching_2024}. These challenges are listed below with the relevant supporting research. While some papers raised additional challenges, these were the five most frequent from our literature search.

\paragraph{The value of ethicists and experts as part of the instruction team.} Researchers have reiterated the challenge of incorporating ethics in a computing-focused course when the instructor or members of the teaching team are not experts in ethics or human-values \cite{ryoo,saltz_integrating_2019,pasricha_ethics_2023,smith_incorporating_2023,tran_its_2024,fiesler_integrating_2021}. In order to incorporate human-centered content effectively and balance this material with established technical learning outcomes, instructors have outlined the need for ethicists to lead certain lectures, bringing their focused expertise.

\paragraph{Difficulties can arise from time constraints, the complexity of embedded material, and its relevance to students.} As ethics and human-centered content is embedded in existing technical courses, there may be an increase in time required for the instruction team to develop this content \textit{and} for students to engage with it on top of developing their technical skills \cite{Padiyath,saltz_integrating_2019,smith_incorporating_2023}. Additionally, the complexity of the embedded material may exceed that of the lecture's computing task (particularly in CS1 courses). Lastly, the ethical content of an embedded course may not be as relevant as the technical components, for those students who do not relate to or anticipate leveraging the ethical learning outcomes. Students may prioritize their grades over advancing their ethical knowledge or decision-making skills, allocating more time to non-ethical components of the course.

\paragraph{Assessments for embedded content may differ from technical skills.} Data science and computer science instructors may have established methods of assessment for technical skills which do not cleanly translate to assessing embedded content
\cite{smith_incorporating_2023}. During course development, evaluation tools may need to be altered, and rubrics developed, to grade students on this additional content. Students and teaching assistants will also need to be informed about how these assessments are graded.

\paragraph{The addition of embedded content can overload the existing curriculum.} One prominent challenge is the overloaded curriculum volume due to the addition of embedded content. If this content is not integrated within the existing material, and instead adds on top of the learning outcomes and lecture material, students may become overwhelmed with the amount of material introduced each week and by the end of the semester \cite{ryoo,smith_incorporating_2023}.


The common challenges to embedding content within technical courses are integrated into the HELIX Framework, as they align with the three pillars. The inclusion of these challenges aims to prepare instructors for any barriers they may face throughout the HCDS education life cycle - within each course and the program as a whole. By acknowledging the potential challenges during course development, they may be mitigated by the instruction team upon first encounter.

\subsubsection{Framework Structure}

The three pillars parallel students' learning and engagement within a course and within a program. Due to the interdisciplinary nature of IS departments and courses, the pillars are structured around actions, learning outcomes, and challenges that are applicable to the wide range of students' educational backgrounds. The framework can be applied to courses from introductory classes in computing, to more advanced technical and special topics subjects. Initiating the framework at knowledge building (remembering, understanding) prompts consideration of the human-centered material that students are instructed through lecture, that they are expected to remember and understand (such as definitions and theories). After encountering and grasping these topics, students move towards decision making, where they apply and analyze these human-centered concepts through activities or reflections. In a data science course, the application and analysis of the prior definitions and theories can include bridging technical and human-centered concepts in assignments and guided activities. Students' decision making processes will differ, and the level to which students uptake and integrate human-centered considerations will be dependent on their educational background, relevant prior work, and experiences in the classroom. Lastly, the framework moves towards empowerment, investing these knowledge and decision making lessons towards future courses and the workplace. This pillar aims to empower students to evaluate and create human-centered technical solutions on their own, bridging their education with future professional careers and continuing their intellectual growth. Behaviours that demonstrate empowerment include students' ability to question and critique existing work practices, raise concerns when applicable, and advocate for responsible design in the workplace. The iterative nature of the framework is important as students' technical skills may not always align with the complexity of the human-centered content they are engaging with, and iterative knowledge building of relevant definitions and theories may be needed before applying these concepts. Likewise, repeated engagement with knowledge building and decision making steps can more adequately prepare students for the workplace and the potential ethical challenges they may face in future careers.

Multiple iterations of the framework were developed, refining the most crucial information to be included on the basis of its practicality for future instructors. It was imperative to consider the educational heterogeneity in students' backgrounds, academic interests, and future career goals, since IS departments prepare students for a wide range of roles. At the graduate level, particularly for introductory computing courses, there is often a lack of prerequisite courses and students come from a wide range of streams. By centering this motivation for the framework, accommodating students across streams and technical abilities, each pillar (Figure \ref{fig:framework}) was visualized to include a set of actions, learning outcomes, and challenges. The circular structure emphasizes the iterative nature of the framework, within a course and also a program, allowing instructors and students to reflect on each pillar as they continue moving forwards to applying these learning outcomes to the professional workplace.

\section{Case Study}
\subsection{Study Setting}

We applied our framework in a preliminary case study as the first iteration of its implementation, taking place at \textcolor{red}{redacted University during the redacted semester}. The goal of the case study was to assess the feasibility of implementing the HELIX Framework in a semester-long course, understand student engagement with the material, and determine if the embedded content had any effect on our students' understanding of human-centered data science. We conducted this case study as the first iteration of our framework's addition to a computing course, and our future work will replicate this study in the same course, as well as other courses in the same stream, to further evaluate the framework's effectiveness. We present our results, including brief quantitative and qualitative analyzes, to provide researchers and educators with an example of how to potentially assess our framework in practice.

The semester-long, graduate-level course offered in an IS department took place once a week for 12 weeks . Ethics approval was obtained by the Research Ethics Board at \textcolor{red}{redacted institution, protocol number XXXXX}. This protocol approved the collection of pre- and post-course surveys from students (given their written consent) 
at the conclusion of the semester. Participation was voluntary, introducing potential self-selection bias as those already interested in ethics-related content may have been more likely to participate. Out of 62 students enrolled in the course, 23 consented to participate (37\%), although one student only completed the pre-course survey and their response was removed for a total of 22 completed pre- and post-course responses. It was made clear that participation in this study had no impact on student grades, and no data was acquired until final grades were approved by the department. There are no specific prerequisites for this introductory programming course or for the overarching IS graduate program itself (other than a prior degree), meaning students come from a variety of educational backgrounds. This course is a requirement for students in the \textcolor{red}{redacted} stream; however, students also took this offering of the course as an elective and/or out of interest. This contributed to a wider background of students from multiple other streams. 

\subsection{Survey Development} \label{sec:survey}
We developed a comprehensive pre- and post-course survey instrument to assess students' understanding of and attitudes toward ethical considerations in data science. The survey followed established practices in computing education research that employ pre- and post-survey designs to measure changes in student knowledge, attitudes, and identity development \cite{jacob_examining_2022}. The instrument was designed to capture both knowledge acquisition and attitudinal changes following exposure to Human-Centered Data Science (HCDS) principles.

The survey consisted of four main sections:

\paragraph{Demographics and Background (2 items):} We collected information about students' prior experience in data science and ethics education to establish baseline knowledge and contextualize responses, following recommendations for computing education survey research that emphasize the importance of understanding learner backgrounds \cite{jacob_examining_2022}.

\paragraph{Knowledge Assessment (6 items):} We developed multiple-choice questions targeting core HCDS concepts including algorithmic bias, data minimization, the socio-political nature of data and algorithms, the ``garbage in, garbage out'' principle from an HCDS perspective, interpretability trade-offs, and limitations of participatory approaches. Questions were designed to assess conceptual understanding rather than rote memorization, with incorrect options reflecting common misconceptions identified in the literature.

\paragraph{Ethical Scenarios (7 items):} Drawing on established practices in computing ethics education that use scenario-based approaches to support ethical decision-making processes \cite{kert_scenarios_2012}, we created scenario-based questions presenting realistic ethical dilemmas in data science contexts, including facial recognition surveillance, biased hiring algorithms, health data consent, discriminatory lending, autonomous vehicle decision-making, workplace monitoring, and smart city tracking. This approach aligns with computing education research demonstrating the effectiveness of scenario-based methods for teaching ethics \cite{brown_teaching_2024}. Each scenario presented four response options representing different ethical frameworks and levels of human-centered thinking. Following best practices for scenario-based assessment that emphasize the value of morally ambiguous situations \cite{norris_investigating_2022}, we deliberately avoided designating ``correct'' answers, instead focusing on capturing students' ethical reasoning processes and priorities.

\paragraph{Attitudes and Perceptions (5 items):} We employed 5-point Likert scales to measure students' confidence in identifying ethical issues, perceived importance of ethical considerations, and understanding of societal impacts, consistent with established survey methodologies in computing education research \cite{jacob_examining_2022}.\\

All students completed a pre- and post-course survey (Appendix \ref{sec:survey_Qs}) in the first and last week of the semester, respectively. The survey was mandatory but ungraded, with students then choosing whether to provide consent for their responses to be used in this study. The surveys were identical except for the addition of three questions (one multiple-choice, two open-ended) regarding students' experiences in the course, which were only included in the post-course survey.

\subsection{Embedded Content} \label{sec:embed}

Students had three hours of contact time each week in the form of one, three-hour class, for 12 weeks. The first 2 hours were dedicated to lecture, and the following hour was dedicated to in-class activities (``tutorials''). The embedded content included readings and in-class activities (Table \ref{tab:activities}), which were assessed through quizzes, assignments, and in-class activity submissions. A human-centered activity - either a reading or an in-class activity - was introduced in most weeks, with a deliverable due that day or the following week. Some in-class activities were adapted from Evan Peck's Ethical Engine repository \cite{ethicalengine}, realigning the activities and reflection questions to complement IS program learning outcomes such as social leadership and sociotechnical impacts of algorithmic design. For example, where an Ethical Engine activity would normally prompt students to consider test cases for their developed algorithm and reflect on potentially-biased effects, we adapted the reflection questions to prompt consideration of downstream implications and social impacts. We additionally had students reflect critically on an improved design approach, potentially involving participatory or community engagement. Table \ref{tab:activities} presents the overarching theme(s) of each embedded activity, broadening the existing Peck activities to more deeply encourage reflection on algorithmic impact, data collection, and morality. This repository has five pre-developed activities where students develop an algorithm for a real-world setting, then reflect on their design choices. These settings included housing assignments, information collection, hiring choices, image manipulation, and disaster relief tasks. 

The readings were pulled from a variety of sources including the ACM's EngageCSEdu Ethics Repository \cite{ethics_repo}, the ACM Digital Library, and government resources on data privacy and regulations \textcolor{red}{[ANON]}. Students reflected on these readings with prompts such as \textit{who is represented in the data, what are the societal implications of this work, and do you have recommendations for future researchers based on this work}? Full marks were allocated for a response that connected with the given reading and made a complete attempt at responding to the prompts. Students also completed a final project during the course, applying a machine learning model to a dataset of their choice and were prompted to discuss societal implications and ethical considerations of their project (this data was not collected for our study).

This structure of the embedded content was chosen in order to complement the natural progression of student learning within a course and the pillars of the HELIX Framework, from knowledge building, to applying this knowledge in case studies  and technical decision making, and ultimately reflecting on their learning (towards empowerment). The complexity of the human-centered activities/assignments increased throughout the semester as students' confidence in identifying and mitigate ethical challenges advanced. The embedded content was integrated alongside technical skill building in lectures and class activities in order to reduce the possibility of overwhelming students with additional course content. We embedded content that met all the PLOs listed in the framework (Figure \ref{fig:framework}), sometimes providing options between readings for students so they could choose a topic of interest to reflect on (Table \ref{tab:activities}). More examples of embedded content and evaluation prompts can be found in the HELIX Materials Kit (\ref{sec:kit}).

\begin{table}[t]
\centering
\resizebox{\textwidth}{!}{
\begin{tabular}{lllll}
\textbf{Format} & \textbf{Content}            & \textbf{Theme}                & \textbf{Assessment}       & \textbf{Week} \\ \hline
Readings      & HCI Design Principles \cite{obrist}        & Critical Theory               & Multiple-Choice Quiz          & 2             \\
Ethical Engine  & Housing Algorithm            & Scoring Algorithm Impacts     & In-Class Activity         & 3             \\
Readings      & Algorithms in Recidivism \cite{berkeley,compas}    & Algorithmic Justice           & Open-Answer Quiz & 4             \\
Readings      & Informed Consent \cite{washington,identity}            & Data Privacy, Informed Consent & Assignment        & 4             \\
Ethical Engine  & Information Collection       & Data Collection               & In-Class Activity         & 5             \\
Ethical Engine  & Hiring Algorithm             & Scoring Algorithm Impacts     & In-Class Activity         & 6             \\
Readings      & Third-Wave HCI  \cite{thirdwave}             & User-Centeredness               & Assignment       & 7             \\
Readings      & Ethics in Data Visualization \cite{vis_ethics} & Ethics                        & Open-Answer Quiz  & 8             \\
Readings      & Fairness in Machine Learning \cite{Obermeyer,gender_shades} & Algorithmic Justice                        & Open-Answer Quiz  & 9             \\
Ethical Engine      & Human Values  & Human Values, Morality                       & In-Class Activity  & 11             \\
                &                              &                               &                           &              
\end{tabular}}
\caption{Case Study Embedded Activities}
\label{tab:activities}
\end{table}

\subsection{Data Analysis} \label{sec:method_quant}

With the goal of preliminarily assessing feasibility of the imparting knowledge when implementing the HELIX Framework in an IS class, participants' pre- and post-course surveys (except the demographics and open-ended responses) were quantitatively compared using Wilcoxon signed-rank tests to assess if there was a statistically significant difference in students' performance on the survey. Each of the knowledge assessment and ethical scenario questions was `scored' from 1 to 5 depending on the human-centeredness of the responses (as guided by prior work in human-centered data science and the Human-Centered Algorithmic Design framework \cite{Chui_Pater_Toscos_Guha_2023,baumer_HCAD,Moon_Guha_2024,McConvey_Guha_Kuzminykh_2023,saxena_lit}, examples provided in Appendix \ref{sec:scoring}). The attitudes and perceptions questions (Likert) were already submitted on a scale of 1 to 5. The Wilcoxon signed-rank test was chosen due to the paired nature of the data, our sample size of 22, and non-normally distributed score differences (between pre- and post-course). Students' scores were accumulated across all questions (knowledge assessment, ethical scenarios, and attitudes and perceptions), separated into pre- and post-course groups, and a Wilcoxon signed-rank test was run to assess for overall change (\ref{sec:quant}).

Students also submitted open-ended responses to two reflection questions. These responses were not extensively thematically analyzed due to their brevity and one student's blank submission leaving only 21 complete responses. Instead, the responses were read and briefly coded by two of the co-authors to count the mentions of different human-centered concepts (ethics, bias, fairness, transparency), see Table \ref{tab:open}. Coder disagreements that arose during this process were resolved through an additional discussion. 


\subsection{Case Study Results} \label{sec:casestudy}


A total of 22 students completed the pre- and post-course surveys and consented to have their responses analyzed for this study. Of those 22, ten had previously studied ethics in a technology/data-focused ethics course, another ten had taken a general ethics course, and two had no experience. When asking students if they had a background in data science or related fields, 15 had completed `some coursework,' two had professional experience, one was self-taught, one was self-taught and had taken coursework, and three had no prior experience.


\subsubsection{Quantitative Analysis of Survey Questions} \label{sec:quant}

The full survey and individual question formats (Likert style, points awarded, open-ended) are listed in Appendix \ref{sec:appendix}).

\begin{table}[t]
\begin{tabular}{lllll}
\textbf{Survey Component}                       & \textbf{Session} & \textbf{Mean} & \textbf{STD} & \textbf{Cohen's d {[}95\% CI{]}} \\ \hline
\multirow{3}{*}{\textbf{Total}}                 & Pre              & 72.3          & 7.0          &                                  \\
& Post             & 75.8          & 4.1          &                                  \\
& Difference       & + 3.5         & 3.9          & -0.61 {[}-1.09, -0.13{]}         \\ \hline
\multirow{3}{*}{\textbf{Knowledge Assessment}}  & Pre              & 26.6          & 2.8          &                                  \\
& Post             & 27.6          & 1.3          &                                  \\
& Difference       & + 0.1         & 1.5          & -0.46 {[}-0.92, 0.01{]}          \\ \hline
\multirow{3}{*}{\textbf{Ethical Scenarios}}     & Pre              & 26.0          & 3.1          &                                  \\
& Post             & 26.3          & 3.3          &                                  \\
& Difference       & + 0.3         & 0.2          & -0.10 {[}-0.54, 0.35{]}          \\ \hline
\multirow{3}{*}{\textbf{Attitudes/Perceptions}} & Pre              & 20.8          & 2.6          &                                  \\
& Post             & 22.7          & 1.7          &                                  \\
& Difference       & + 1.9         & 0.9          & -0.88 {[}-1.40, -0.36{]}        
\end{tabular}
\caption{Descriptive Quantitative Survey Results}
\label{tab:diffs}
\end{table}

There was a statistically significant difference in students' pre- and post-course responses to these five Likert questions (p = 0.002). The average pre-course Likert score was 20.8, while the post-course average was 22.7. 

As reflected in Table \ref{tab:diffs}, the scores on each section of the survey, as well as the overall scores, increased in mean between pre- and post-course completion. As well, the standard deviations of student responses were lower in all components, post-course (except ethical scenarios, which saw the least change overall). The Cohen's d statistic provides a quantitative description of the difference in distributions for student responses in each category. A power analysis was conducted for the sample size of 22, resulting in a power of 0.61. To increase the power to above 0.8, a sample size of 34 would be required, reinforcing the appropriateness of the Wilcoxon signed-rank test for our study.


\begin{table}[t]
\begin{tabular}{ll|l|l}
           &                                      & \textbf{Workplace Challenges} & \textbf{Approach to Decision Making} \\ \hline
\multicolumn{2}{l|}{\textbf{Complete Responses}}  & 21                                    & 21                                           \\
           & More Human-Centered                  & 21                                    & 19                                           \\
           & Neutral                              & 0                                     & 2   \\
           & Less Human-Centered
           & 0
           & 0 \\ \hline
\multicolumn{2}{l|}{\textbf{Considerations Noted}} &                                       &                                              \\
           & Ethics                               & 11                                    & 14                                           \\
           & Privacy                              & 15                                    & 4                                            \\
           & Bias/Fairness                        & 8                                     & 9                                           \\
           & Societal Impact                      & 6                                     & 12                                           \\
           & Transparency/Explainability          & 3                                     & 5                                            \\
           & Developer Accountability             & 0                                     & 3                                           
\end{tabular}
\caption{Tabulated Responses to Open-Ended Survey Questions}
\label{tab:open}
\end{table}

In the post-course survey only, students were asked: \textit{Compared to before the course, how has your understanding of ethical data science changed?}. Out of 22 respondents, 18 selected `somewhat improved,' 3 selected `significantly improved,' while only 1 felt `no change.' Students were also asked to respond to the following open-ended question: \textit{What ethical challenges do you anticipate facing in your future data science work?} The `Workplace Challenges' column in Table \ref{tab:open} reflects those considerations students mentioned in their responses, N = 21 (one response was left blank). Privacy concerns were noted most frequently (71.4\% respondents), with ethical concerns (52.4\%) and bias and fairness concerns (38.1\%) following. The final question of the post-course survey asked students to submit an open-ended response to the following question: \textit{How has this course influenced your approach to ethical decision-making in data science?} The `Approach to Decision Making' column in Table \ref{tab:open} shows those considerations mentioned in 21 responses. 14 respondents identified an increased need for conscious ethical decision making (66.7\%), with 12 emphasizing societal impacts of their work (57.1\%), and 9 prioritizing bias and fairness in the workplace (42.9\%).


Students' scores across all questions, including all sections except the open-ended responses and demographics, were summed, and a Wilcoxon signed-rank test (N = 22) was conducted to assess for a statistically significant different in total score. This result was significant, with p = 0.001. Our case study indicates a \textbf{significant difference in students' performance on this survey when taken pre- and post-course}. This suggests the integration of our {HELIX Framework may have increased students' knowledge gains regarding human-centered, societal impact, justice, bias, and ethical content. To assess the efficacy of the HELIX Framework, future studies will be conducted, specifically in courses that teach technical topics more advanced than introductory programming.

\begin{figure}[t]
    \centering
    \includegraphics[width=0.5\linewidth]{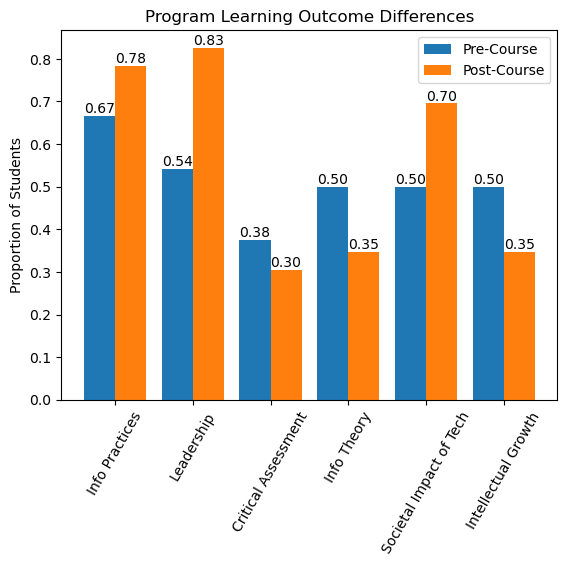}
    \caption{Anticipated and Advanced PLOs. From left to right, the following PLO labels are briefly defined. `Info Practices:' understanding information theories, practices, and disciplines; adapting to needs of society. `Leadership:' future leadership skills and social responsibility to provide information services for all. `Critical Assessment:' contributing to the expansion of the information science field through critical assessment and research. `Info Theory:' understanding the development of theory concerning information. `Societal Impact of Tech:' understanding new technological developments as they relate to information preservation; identifying societal impacts. `Intellectual Growth:' life-long learning.}
    \label{fig:PLOs}
\end{figure}

\subsubsection{Program Learning Outcomes}

Students also reflected on the six PLOs specific to their graduate degree, as defined by the department. These PLOs cover a range of learning goals, from advancing skills in social leadership and critical assessment, to understanding new technologies, their societal impacts, and evolving information theories. 
In the pre-course survey, students selected all the PLOs they expected to advance during the course. They then selected those PLOs they felt they did advance in the post-course survey. Figure \ref{fig:PLOs} compares the proportion of students that anticipated (pre-course) and advanced (post-course) each PLO, with a brief description of the PLO.

There was an increase in students' learning regarding information practices and social leadership, alongside understanding the societal impacts of technology. However, a higher number of students anticipated learning about critical assessment and information theory, compared with post-course responses. The final PLO relates to future intellectual growth and continuous learning, which students again felt would be advanced more at the beginning of the semester. Not all PLOs will be directly related to the content of every course in an IS department, meaning further iterations of our HELIX Framework application can try to accommodate those PLOs that students felt they did not advance in this iteration, while acknowledging the benefit of other courses in their IS degrees to complement this learning.

\subsection{HELIX Materials Kit} \label{sec:kit}
We developed the HELIX Materials Kit v1.0 \cite{zenodo_id}
for researchers' and educators' future reference and use. The kit includes our complete survey and each question's scoring rubric, our adapted Ethical Engine assignments, evaluated activity and quiz prompts, a sample dataset with synthetic survey responses, and a code notebook to quantitatively analyze the sample dataset as described in \ref{sec:method_quant}. 

\section{Discussion}

\subsection{Implications of HELIX Framework Case Study and Potential Improvements}

The quantitative results from our single, preliminary case study (see \ref{sec:casestudy}), alongside the coded open-ended responses and changes in PLOs advanced suggest the HELIX Framework is applicable to introductory programming courses in an IS department. There was a statistically significant difference (p < 0.01) in students' overall performance on the knowledge assessment, ethical scenario, and attitudes and perceptions questions, between the first and last weeks of the semester. This suggests the implementation of human-centered content (Table \ref{tab:activities}) throughout the semester increased students' performance on our survey questions and the human-centered decision making we prompted students to consider. The survey was developed around the three pillars of the framework (\ref{sec:survey}, Figure \ref{fig:framework}), beginning with knowledge assessment, moving into ethical decision making, and finally towards attitudes and perceptions (concluding with open-ended reflections in the post-course survey). While attitudes toward human-centered responsibility improved, our initial case study found minimal change in measured knowledge and scenario-based decision making. This suggests that short-term exposure may be sufficient to shift awareness, but insufficient to develop applied ethical reasoning skills. Future studies will explore further implementation of the HELIX framework in data-focused courses, to understand how human-centered lessons can more deeply embed responsible design principles beyond the classroom. From this case study, we summarize three takeaways for future implementation of this framework in the same setting (introductory programming course in an IS department):

\begin{enumerate}
    \item Knowledge building activities must satisfy a broad range of student interests. Students from IS departments may come from a range of disciplines within the degree. To complement their academic and educational interests, a range of datasets, real-world examples, and take-home activities can increase their engagement with human-centered concepts (which may be entirely new to some students). In our case study, we utilized a broad range of examples and datasets (Table \ref{tab:activities}), with the ultimate goal of connecting students to the data and providing tangible, achievable goals for their developing technical skills to achieve. Adapting knowledge building content to your students' interests can increase their engagement with human-centered thinking early in the course, propelling their advancement through decision making and empowerment pillars.
    \item Embedding hands-on activities can increase accountability when designing algorithmic solutions and alter perceptions of relevant ethical challenges. From the results of our survey and the open-ended reflections, our students have an increased awareness of ethics, privacy, bias, fairness, societal impact, transparency, accountability, and critical thinking (Table \ref{tab:open}) regarding technological developments and their role as developers. The inclusion of ethical design activities during four of the in-class activities provided students with tangible examples and references regarding ethical decision making for data-driven solutions in practice. These activities also prompt students to take ownership of their skill development and decision making, encouraging active learning through real-world examples. Developing these assignments to bring students into human-centered collaboration with instructors provides them with greater autonomy over their critical thinking and hones their coding intuition, with previous efforts encouraging active learning in data science at multiple stages of education \cite{active_learning_1,sanusi}.
    \item Prompting reflection on societal impact of one's algorithmic design can increase students' critiquing abilities for technological developments, but this does not always transfer to the workplace. Our adapted Ethical Engine activities prompted students to reflect on the impacts of their algorithmic developments; however, for students from non-technical backgrounds or with primarily non-technical academic interests, it was difficult to contextualize how these skills might apply to their future roles as information professionals. To increase the applicability of these skills and the learning outcomes, a broader range of activities should be incorporated to include applications of data in those non-technical roles.
\end{enumerate}

We also highlight the need for nuance regarding empowerment of students moving beyond a particular course. The application of this framework may not fully provide students with the ability and confidence to contest unethical decisions in the workplace (\ref{sec:challenges}), and should instead focus on empowering students to recognize these tensions and feel motivated to seek additional resources and solutions. As this is the first iteration of the implementation of the HELIX Framework, further research into the application of this framework in introductory programming courses, as well as more technically advanced courses, is necessary to validate these findings. We encourage researchers to apply the HELIX Framework with their students, see additional guidelines in \ref{sec:guidelines}, and to assess students' receptiveness to the embedded content in any relevant manner.

\subsection{Implications of the HELIX Framework for Computing Instruction in IS}

The development of the HELIX Framework comes at a time when data science continues proliferating in post-secondary education, ranging  from departments such as computer science to statistics, IS, public health, and more \cite{Padiyath}. As data science programs grow out of the need for experts in data-driven solutions, they must adapt to the educational settings of the departments they reside in. Prior SIGCHI research has emphasized the need for ethics education in computer science courses, to encourage students to make conscious, human-centered choices and reduce harm with their future academic and workplace algorithmic developments \cite{horton_embedded_2024,saltz_integrating_2019,skirpan_ethics_2018,Hedayati}. This must be mirrored in departments outside of computer science, that also aim to prepare students for professional careers in data but include learning outcomes outside of a mostly technical domain. We fill this niche in IS departments by developing the HELIX Framework from an extensive literature scan of curricular frameworks, widespread IS PLOs, and well known challenges in embedding ethics (\ref{sec:framework}). This framework is novel in its setting, intended primarily for IS departments that provide instruction on data science and welcome human-centered content and solutions. Lastly, the framework prompts interesting considerations of how human-centered content (ethics, justice, bias, fairness, morality, critical theory, privacy, consent, algorithmic impact, user-centeredness, etc.) can be integrated alongside technical skill building - from beginner computing courses to advanced, specialized topics. This integration may require consultation from experts and ethicists, to fully inform students of the implications of their work from a non-technical perspective.


The development of HELIX was motivated by emerging trends in IS and computing science course learning outcomes, which increasingly emphasize critical reflection on the societal impacts of data-driven systems, responsible decision-making, and learner agency.
Our coding of PLOs across iSchools reveals limited explicit attention to human-centered principles, ethics, leadership, and/or social justice. This finding is notable given that human-centeredness and social responsibility have been identified as distinctive features of the iSchool approach to data science \cite{shah}.
iSchools would frequently list 1+ course(s) centered on responsible or ethical decision making; however, these `one-off' course offerings would not be reflected in the higher-level themes of the program which instead most frequently focused on information practices, technological developments, and communication. AI, data science, and/or Big Data streams are becoming more popular within these iSchool programs---both graduate and undergraduate---yet the overall PLOs do not reflect the goals of producing reflexive, justice-centered professionals. As a result, instructors often lack a coherent structure for emphasizing the integration of human-centered learning into technical course offerings. HELIX addresses this gap by organizing data science learning around the interconnected and human-centered pillars of knowledge building, decision-making, and empowerment. For instructors, the framework provides actionable guidance for designing activities and assessments that integrate societal considerations without displacing technical content. For students, it offers opportunities to connect programming and computational concepts with real-world consequences, fostering both awareness and a sense of agency. As conducted in our case study, at the classroom level, HELIX serves as a scaffold for embedding human-centered perspectives throughout a course rather than treating them as isolated topics, supporting a more holistic approach to human-centered computing education while mitigating anticipated challenges from prior embedded ethics studies.


Applying the HELIX Framework does not require replacing existing technical learning outcomes but reframes and extends them through a human-centered lens. In the introductory programming course examined in our preliminary case study, activities were designed and adapted to complement core objectives related to coding, problem solving, and computational thinking by encouraging students to consider how technical decisions shape user experiences and societal outcomes (Table \ref{tab:open}). This approach can enrich expected technical learning by connecting abstract computing concepts to authentic contexts and fostering broader professional competencies (as reflected in the open-ended responses and PLO reflections - Table \ref{tab:open}, Figure \ref{fig:PLOs}). At the same time, integrating human-centered content introduces several design tensions. Additional readings, reflections, and discussion activities may compete for limited instructional time, creating concerns about workload for both students and instructors. Assessing growth in attitudes, ethical reasoning, or empowerment can also be more challenging than evaluating technical skills. Furthermore, instructors must balance maintaining relevance for learners with diverse career goals and educational backgrounds while ensuring that human-centered topics are meaningfully integrated rather than treated as ancillary material. These considerations highlight both the potential benefits and the practical trade-offs associated with implementing HELIX in technically focused courses. Should an IS department implement this framework throughout their data science degrees, students will receive an education built around a broader range of learning outcomes. Not only are students' technical skills expected to advance, alongside their understanding of relevant human-centered issues, but their ability to integrate these considerations in their future work and algorithmic design tasks may advance. The bridging of technical and theoretical can provide additional competencies than expected for primarily technical courses.

\subsection{Guidelines for the Application of the HELIX Framework in Practice} \label{sec:guidelines}

We provide guidelines for future researchers and educators to consider and apply when utilizing the HELIX Framework in practice. These guidelines are rooted in our case study experience as well as the prior challenges highlighted in embedded ethics research.

\begin{enumerate}
    \item \textbf{Encourage flexibility when applying the HELIX Framework, allowing students to control the flow between pillars.} Particularly for students from a technical background, who have not engaged with human-centered material before, the rate at which these topics will be understood, analyzed, and evaluated will differ from a purely technical course. Remembering that the pillars of the framework are iterative, both within a single course and across a program, can help students deeply engage with new concepts before moving towards decision making and empowerment. 
    \item \textbf{Think ahead to future courses for your students and introduce human-centered concepts that complement those curricula.} Providing continuous ethical and human-centered lessons across students' programs can yield a prolonged, cohesive, and rich  toolbox for ethical choices in the workplace, and encourage students to feel empowered when facing challenges or barriers in their decision making \cite{horton_embedded_2024}. For students in our case study, those in the \textcolor{red}{redacted} stream are required to take a future course on human values and its intersection with data science. By increasing students' understanding and comfort with the concepts introduced in that course (usually the following year), future engagement with human-centered ideas may increase.
    \item \textbf{Integrate the human-centered content with the technical skill building aspects of the course}. One common challenge in embedded ethics courses is the increase in course volume \cite{ryoo,smith_incorporating_2023,brown_teaching_2024}. Students have reported feeling overwhelmed in such courses due to the additional content. By instead integrating the embedded content with technical applications and computing examples, students can engage with real-world topics from multiple perspectives. This can combat the barrier of an overwhelming course load, as reported by our case study's students in their open-ended responses - who appreciated the use of technical examples and datasets to explore ethical challenges.
    \item \textbf{Recognize that embedded content may empower students' decision making in future computing courses; however, bringing these considerations to the workplace remain difficult.} Echoing Horton et al.'s call for continuous training for entry-level workers in data-driven roles, we highlight the difficulty of bringing the lessons in ethical decision making to a new workplace \cite{horton_embedded_2024}. Our case study's open-ended responses included reflections on eagerness to bring this content to the workplace as well as realistic barriers that may prevent them from fully utilizing these tools (specifically conflicts of interest or superiors with differing opinions). We encourage educators to reflect on this challenge and provide students with resources that they may return to in a workplace setting (such as online readings, case studies, or governing bodies in their location). 
\end{enumerate}

Incorporating human-centered content in a computing course will always be a unique opportunity and challenge. By understanding your students' educational and work backgrounds, their future courses with the program, and their background understanding of ethical decision making, the HELIX Framework and Materials Kit can be more meaningfully adapted and integrated.



\section{Conclusion}

As computational programs in data, predominantly including data science, continue increasing throughout non-computer science departments, the need for human-centered, ethical instruction is important. To complement the technical skills being introduced in data science courses, embedded instruction in algorithmic justice, bias, societal impact, and morality is necessary and timely when training the next generation of professionals in data-driven solutions. We developed the Human-Centered Education for Learning in Information and eXplainable Computing (HELIX) Framework to help instructional teams in information science incorporate human-centered content and pursue parallel learning outcomes throughout a single course and an entire degree. HELIX follows iterative stages of students' development in secondary education, including knowledge building, decision making, and empowerment steps, helping instructors propel data science professionals towards ethical decision making. 

We conducted a single case study to test the framework in \textcolor{red}{redacted semester}, teaching introductory programming skills to graduate students, and note two limitations when applying the framework. First, since students in our course had a range of educational backgrounds and are pursuing a range of streams within the broader degree, many took the course out of interest and found it difficult to anticipate the effects of the embedded content in their future workplace. These students struggled to picture the impacts of the human-centered learning outcomes on their future work and decision making. Second, students were required to opt-in to the study and were not automatically enrolled, meaning that those with the most confidence in, and comfort with, these concepts chose to opt in, as well as a lack of control group. Those students who may have learned the most may also not have given consent for their data to be used. Results suggest improvements in students' attitudes toward ethical responsibility, while knowledge and decision-making skills showed minimal change. These findings indicate the HELIX Framework's potential as a pedagogical instrument in interdisciplinary computing contexts, but highlight the need for more rigorous evaluation.

Our framework aims to inspire embedded human-centered content in data science courses, to further entwine students' reflection skills with their technical abilities. To broaden students' critical thinking and ethical decision making abilities, we encourage researchers and educators to adapt and apply the Human-Centered Education for Learning in Information and eXplainable Computing (HELIX) Framework, encouraging more ethical data science learning outcomes and decision making in the workplace.

\bibliographystyle{ACM-Reference-Format}
\bibliography{pedagogy,references}

\appendix
\section{Appendix} \label{sec:appendix}

\subsection{Survey Questions} \label{sec:survey_Qs}

\subsubsection{Demographics}
\begin{enumerate}
    \item What is your background in data science or related fields? (Select all that apply)
    \begin{itemize}
        \item No prior experience
        \item Some coursework
        \item Professional experience
        \item Self-taught
    \end{itemize}
    \item Have you previously studied ethics in any context?
    \begin{itemize}
        \item No
        \item Yes, in a general ethics course
        \item Yes, in a technology/data-focused ethics course
    \end{itemize}
\end{enumerate}

\subsubsection{Knowledge Assessment}
\begin{enumerate}
    \item The concept of 'algorithmic bias' refers to:
    \begin{itemize}
        \item A mathematical error in a dataset 
        \item A systematic and unfair disadvantage for certain groups due to how an algorithm is trained or deployed
        \item A preference for certain data types in machine learning models
        \item A random error that occurs during data processing
    \end{itemize}

    \item The term 'data minimization' means:
    \begin{itemize}
        \item Collecting as much data as possible to improve accuracy
        \item Keeping only the necessary amount of data for a specific purpose
        \item Sharing data freely across organizations for innovation
        \item Reducing dataset size to make computation faster
    \end{itemize}

    \item Human-Centered Data Science (HCDS) challenges the assumption that data and algorithms are inherently neutral. Which of the following best explains why this assumption is flawed?
    \begin{itemize}
        \item The bias in algorithms stems solely from mathematical errors.
        \item Data and algorithms are shaped by the social, historical, and political contexts in which they are created.
        \item Algorithmic bias is an unavoidable consequence of data science and cannot be mitigated or fixed.
        \item A well-designed algorithm can eliminate bias.
    \end{itemize}

     \item The "garbage in, garbage out" principle in machine learning is a well-known problem. From an HCDS perspective, what additional challenges does this principle introduce?
    \begin{itemize}
        \item Even high-quality datasets may encode systemic inequalities that are difficult to detect using standard evaluation metrics.
        \item The problem is purely technical and can be fixed by increasing dataset size.
        \item High-performing models are inherently resistant to biased input data.
        \item Algorithmic bias can only occur if there are explicit discriminatory labels in the dataset.
    \end{itemize}

    \item Human-Centered Data Science emphasizes interpretability and explainability in AI models. Why might this be a double-edged sword?
    \begin{itemize}
        \item Increasing model transparency may inadvertently reveal proprietary trade secrets, limiting adoption by companies.
        \item More interpretable models tend to have higher computational costs, making them less scalable.
        \item Explainable models may give users a false sense of trust, as simplified explanations can mask deeper ethical concerns.
        \item All of the above.
    \end{itemize}

     \item What is a limitation of participatory approaches in Human-Centered Data Science?
    \begin{itemize}
        \item Including diverse stakeholders can increase project complexity and require longer development cycles.
        \item Non-experts should not be involved in shaping AI and data science projects.
        \item Participation in Ai development necessarily results in better ethical outcomes.
        \item Stakeholder input is only valuable during the deployment phase, not during model development.
    \end{itemize}
\end{enumerate}

\subsubsection{Ethical Scenarios} The following section presents seven data-related ethical scenarios. There are no right or wrong answers. Please pick the answer with which you most agree.

\begin{enumerate}
    
    \item You are part of a research team working on an AI-powered facial recognition designed to prevent violent crimes in public spaces. The system improves public safety but raises concerns about privacy and potential misuse by law enforcement.

    Which perspective best aligns with your current thinking?
    \begin{itemize}
        \item Public safety comes first---if the system can reduce crime, the benefits outweigh privacy concerns. 
        \item Privacy must be prioritized---mass surveillance could set a dangerous precedent, even if crime prevention is improved. 
        \item A middle ground is necessary---strict oversight and transparency should be required to prevent misuse while allowing some safety benefits.
        \item Technology should not make this decision---these systems should not be deployed unless broader social policies address root causes of crime.
    \end{itemize}

    \item An AI-driven hiring tool predicts candidate success based on past employee data. However, an audit reveals that it favors applicants from prestigious universities and filters out those from non-traditional backgrounds.
    What is the best way to handle this issue?

    \begin{itemize}
        \item Let the algorithm stand—it reflects real-world hiring trends, and companies should prioritize efficiency.

        \item Adjust the algorithm—incorporate fairness constraints to ensure a more diverse pool of candidates.

        \item Intervene outside the model—companies should change their hiring policies rather than modifying the AI.

        \item Abolish AI hiring tools—human decision-making is necessary to prevent structural discrimination.

    \end{itemize}

    \item A startup’s AI model predicts disease outbreaks using anonymized health and mobility data, but users were not informed that their data was collected.
    What is the most responsible approach?

    \begin{itemize}
        \item Use the data as is—public health benefits outweigh concerns about individual consent.

        \item Introduce post hoc consent—inform users now and allow them to opt out moving forward.
        \item Strengthen privacy protections—implement differential privacy techniques to balance utility and ethics.

        \item Discard the dataset—if users did not consent initially, the data should not be used.

    \end{itemize}

   \item A bank’s AI-driven loan approval system disproportionately denies loans to low-income applicants due to historical lending patterns.
   What should be done?

    \begin{itemize}
        \item Maintain the system—the AI is objective, and financial risk should not be artificially adjusted.

        \item Modify the model—adjust approval criteria to account for systemic inequities.

        \item Introduce alternative lending policies—use human decision-makers alongside AI recommendations.

        \item Pause AI use in lending—until bias is fully understood, loans should not be determined algorithmically.

    \end{itemize}

    \item  During testing, an AI-powered self-driving car must choose between swerving to protect passengers or staying its course, potentially harming pedestrians.
    Which principle should guide decision-making?

    \begin{itemize}
        \item Passenger safety first—the car’s primary duty is to protect its occupants.
        
        \item Minimize total harm—whichever decision results in fewer casualties should be chosen.
        
        \item Ethical customization—users should be able to set ethical preferences for their vehicles.
        
        \item Regulatory decisions—governments, not AI developers, should define the rules for these situations.
    \end{itemize}

    \item  A company deploys AI to track employee productivity via keystroke monitoring and email analysis. Employees feel this is invasive, but management claims it prevents burnout and increases efficiency.
    What is the most justifiable stance?

    \begin{itemize}
        \item AI monitoring improves productivity—as long as data is used responsibly, the benefits outweigh concerns.

        \item Strict regulations needed—monitoring should be transparent, and employees must give informed consent.

        \item Employee autonomy matters—AI should not be used for workplace surveillance at all.

        \item Industry-dependent approach—highly regulated sectors (e.g., finance, healthcare) may justify more oversight than others.

    \end{itemize}

    \item A city plans to deploy AI-powered infrastructure that optimizes traffic, energy use, and pollution reduction, but also tracks citizen movements.
    What approach should be taken?

    \begin{itemize}
        \item Full deployment—the efficiency benefits outweigh privacy concerns.
 
        \item Limited use with opt-out options—citizens should have a say in how their data is used.
        
        \item Independent oversight—an external ethics board should regulate the project.
       
        \item Reject mass surveillance—urban planning should not depend on citizen tracking.
    \end{itemize}
\end{enumerate}

\subsubsection{Attitudes and Perceptions} 
\begin{enumerate} 

    \item How strongly do you agree with the following statements? \\
    \textit{Rate from 1 to 5: 1 = Strong Disagree, 5 = Strong Agree}
    \begin{itemize}
        \item (a) Ethical considerations should be a fundamental part of data science.

        \item (b) I feel confident in identifying ethical issues in data science.

        \item (c) Data privacy is an important concern in my field of study.

        \item (d) Companies should be held accountable for biased AI systems.

        \item (e) I understand the societal impact of biased data and algorithms.

    \end{itemize}
\end{enumerate}

\subsubsection{Post-Course Only: Course Reflection Questions}
\begin{enumerate}
    \item Compared to before the course, how has your understanding of ethical data science changed?
    \begin{itemize}
        \item No change
        \item Somewhat improved
        \item Significantly improved
    \end{itemize}
    \item What ethical challenges do you anticipate facing in your future data science work? \textit{(Open-ended response)}
    \item How has this course influenced your approach to ethical decision-making in data science? \textit{(Open-ended response)}
    
\end{enumerate}  

\subsection{Scoring Rubric and Examples for Ethical Scenarios} \label{sec:scoring}

Each response was scored from 1-5 based on human-centered orientation:

\begin{itemize}
    \item 5 = Most Human-Centered: Prioritizes human agency, consent, privacy, social justice
    \item 4 = Highly Human-Centered: Balances human concerns with technical solutions
    \item 3 = Moderately Human-Centered: Seeks middle ground or oversight
    \item 2 = Minimally Human-Centered: Some human consideration but efficiency-focused
    \item 1 = Least Human-Centered: Purely utilitarian, ignores human impact
\end{itemize}

Two examples are provided for HCDS scoring of the ethical scenario questions. Additional score rubrics can be found in the HELIX Materials Kit v1.0 \cite{zenodo_id}. Scores for each response are given in \textcolor{blue}{blue text}.

\begin{enumerate}
    \item You are part of a research team working on an AI=powered facial recognition designed to prevent violent crimes in public spaces. The system improves public safety but raises concerns about privacy and potential misuse by law enforcement.

Which perspective best aligns with your current thinking?
    \begin{itemize}
        \item Public safety comes first---if the system can reduce crime, the benefits outweigh privacy concerns. \textcolor{blue}{(1)}
        \item Privacy must be prioritized---mass surveillance could set a dangerous precedent, even if crime prevention is improved. \textcolor{blue}{(5)} 
        \item A middle ground is necessary---strict oversight and transparency should be required to prevent misuse while allowing some safety benefits. \textcolor{blue}{(3)}
        \item Technology should not make this decision---these systems should not be deployed unless broader social policies address root causes of crime. \textcolor{blue}{(4)}
    \end{itemize}
    \item An AI-driven hiring tool predicts candidate success based on past employee data. However, an audit reveals that it favors applicants from prestigious universities and filters out those from non-traditional backgrounds.
What is the best way to handle this issue?

    \begin{itemize}
        \item Let the algorithm stand—it reflects real-world hiring trends, and companies should prioritize efficiency. \textcolor{blue}{(1)}

        \item Adjust the algorithm—incorporate fairness constraints to ensure a more diverse pool of candidates. \textcolor{blue}{(4)}

        \item Intervene outside the model—companies should change their hiring policies rather than modifying the AI. \textcolor{blue}{(3)}

        \item Abolish AI hiring tools—human decision-making is necessary to prevent structural discrimination. \textcolor{blue}{(5)}

    \end{itemize}
\end{enumerate}

\end{document}